# ANALISIS KOPLING MEDAN ELEKTROMAGNETIK *TRANSVERSE MAGNETIC* (TM) PADA KRISTAL FOTONIK 2D DENGAN DEFEK INDEKS BIAS SIMETRIK MENGGUNAKAN METODE TENSOR GREEN

**CANDRA KURNIAWAN**

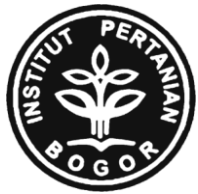

**DEPARTEMEN FISIKA**
**FAKULTAS MATEMATIKA DAN ILMU PENGETAHUAN ALAM**
**INSTITUT PERTANIAN BOGOR**
**2010**

# ABSTRAK

**Candra Kurniawan.** Analisis Kopling Medan Elektromagnetik *Transverse Magnetic* (TM) pada Kristal Fotonik 2D dengan Defek Indeks Bias Simetrik Menggunakan Metode Tensor Green. Dibimbing oleh **Hendradi Hardhienata, M.Si** dan **Dr. Husin Alatas, M.Si**.

Kristal fotonik adalah material dielektrik yang memiliki indeks bias atau permitivitas berbeda secara periodik, sehingga dapat mencegah perambatan cahaya dengan frekuensi dan arah tertentu. Rentang daerah frekuensi tersebut dinamakan *photonic bandgap* (PBG). Jika struktur Kristal fotonik dimodifikasi dengan mengambil satu baris silinder (*rod*) dalam kristal fotonik maka didapatkan sebuah pandu gelombang (*waveguide*). Dengan mamberikan struktur defek (cacat) simetris di sekitar kanal pandu gelombang dengan parameter-parameter kristal tertentu maka terjadi kopling antara kanal dan defek tersebut. Kopling yang terjadi menandakan pengalihan sebagian atau keseluruhan medan EM yang dirambatkan tergantung pada besarnya frekuensi yang dirambatkan dalam kristal fotonik tersebut. Metode tensor green dapat digunakan untuk menghitung kuat medan listrik total dalam suatu krital fotonik 2D. Dengan bantuan program MATLAB maka dapat ditunjukkan visualisasi penghitungan kuat medan listrik total pada kristal fotonik 2D. Melalui analisis grafik *bandgap* dari literatur serta grafik energi medan pada defek pada penelitian ini terlihat bahwa frekuensi efektif yang menghasilkan energi maksimum pada defek pada saat terjadi kopling berada di pertengahan selang frekuensi *bandgap*.

Kata kunci : Kristal fotonik, *photonic bandgap*, pandu gelombang, defek, kopling.

Judul Skripsi : Analisis Kopling Medan Elektromagnetik *Transverse Magnetic* (TM) pada Kristal Fotonik 2D dengan Defek Indeks Bias Simetrik Menggunakan Metode Tensor Green
Nama : Candra Kurniawan
NIM : G74060034

Menyetujui,

Pembimbing I,          Pembimbing II,

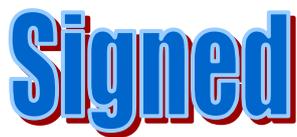          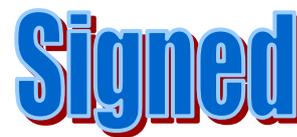

(Hendradi Hardhienata, M.Si)          (Dr. Husin Alatas, M. Si)

NIP: 19830114 200812 1 001          NIP: 19710604 199802 1 001

Mengetahui:

Ketua Departemen Fisika

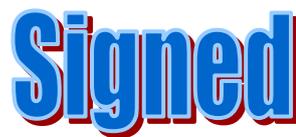

(Dr. Ir. Irzaman, M.Si)

NIP: 19630708 199512 1 001

Tanggal Lulus : **12 Maret 2010**

# ANALISIS KOPLING MEDAN ELEKTROMAGNETIK *TRANSVERSE MAGNETIK* (TM) PADA KRISTAL FOTONIK 2D DENGAN DEFEK INDEKS BIAS SIMETRIK MENGGUNAKAN METODE TENSOR GREEN

Skripsi

Sebagai salah satu syarat untuk memperoleh gelar Sarjana Sains
pada Fakultas Matematika dan Ilmu Pengetahuan Alam
Institut Pertanian Bogor

Oleh:

**CANDRA KURNIAWAN**

**G74060034**

**DEPARTEMEN FISIKA**
**FAKULTAS MATEMATIKA DAN ILMU PENGETAHUAN ALAM**
**INSTITUT PERTANIAN BOGOR**
**2010**

# RIWAYAT HIDUP

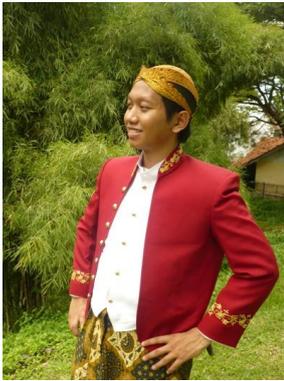

Penulis dilahirkan di DKI Jakarta pada tanggal 3 Oktober 1988. Penulis adalah anak pertama dari dua bersaudara, dari pasangan Bapak Kateman dan Ibu Sari Hartanti.

Pendidikan formal yang penulis tempuh yaitu SD Kartika X-4 Jakarta dan lulus tahun 2000, SLTPN 177 Jakarta dan lulus pada tahun 2003, dan SMUN 47 Jakarta dan lulus pada tahun 2006. Pada tahun yang sama penulis langsung melanjutkan ke perguruan tinggi dan diterima di Institut Pertanian Bogor melalui jalur USMI (Undangan Seleksi Masuk IPB). Pada tahun pertama di IPB penulis menempuh pendidikan Tingkat Persiapan Bersama dan pada tahun 2007 penulis diterima masuk dalam Mayor Fisika, Fakultas Matematika dan Ilmu Pegetahuan Alam. Selama masa pendidikan di Fisika penulis mengambil mata kulah-mata kuliah *Suporting Course* dari beberapa Departemen, diantaranya dari Matematika (kalkulus II dan kalkulus III), Geofisika dan Meteorologi (Meteorologi Fisis), dan Manajemen (Manajemen Pemasaran dan Manajemen Produksi dan Operasi).

Selama masa perkuliahan penulis juga aktif dalam berbagai kegiatan intrakampus maupun ekstrakampus. Penulis pernah menjadi anggota divisi SDM Himpunan Mahasiswa Fisika (2007/2008), Badan Pengawas Bidang Keilmuan Himpunan Mahasiswa Fisika (2009) dan Anggota Ikatan Alumni SMA Sepesanggrahan dan Sekitarnya (IAS3). Penulis pernah mengikuti beberapa kepanitiaan yaitu, sebagai tim khusus Pesta Sains-Kompetisi Fisika (2007), sebagai logistik dan transportasi G-Force MIPA (2008), sebagai tim khusus pusat Pesta Sains (2008), sebagai acara G-Action MIPA (2008), ketua panitia Fieldtrip Fisika "Φ-Trip" (2009), koordinator acara Upgrading HIMAFI (2009), koordinator acara *open house* departemen Fisika (2009), dan sebagai komisi disiplin Masa Perkenalan Departemen Fisika (2009). Penulis juga ikut dalam kegiatan-kegiatan lainnya yaitu sebagai asisten praktikum Fisika TPB (2008-2010), sebagai asisten dosen Fisika Modern (2009) dan Eksperimen Fisika 1 (2009), dan sebagai pengajar Fisika pada bimbingan belajar BINTANG PELAJAR (2009-sekarang).

# KATA PENGANTAR

*Assalamu'alaikum warahmatullahi wabarakatuh.*

*Alhamdulillahirrabbil'alamiin*. Segala puji hanya bagi Allah, *Rab* Semesta Alam. Atas rahmat dan karunia –Nya penulis dapat menyusun skripsi ini yang berjudul "Analisis Kopling Medan Elektromagnetik *Transverse Magnetic* (TM) pada Kristal Fotonik 2D dengan Defek Indeks Bias Simetrik Menggunakan Metode Tensor Green".

Ucapan terima kasih penulis sampaikan pada beberapa pihak yang telah menjadi bagian hidup penulis dan membantu dalam menyusun penelitian ini :

1. Bapak Hendradi Hardhienata, M.Si sebagai pembimbing utama penulis dalam penelitian ini atas nasehat-nasehat dan fasilitas yang telah diberikan kepada penulis, dan Bapak Dr. Husin Alatas, M.Si sebagai pembimbing II serta sebagai kepala bagian fisika teori yang telah memberikan kesempatan pada penulis untuk bergabung dalam keluarga fisika teori IPB dari mulai penulis berada di Departemen Fisika IPB, dan atas segala nasehat yang diberikan kepada penulis.
2. Bapak Dr. Ir.Irzaman, M.Si sebagai ketua Departemen Fisika IPB yang telah memberikan banyak nasehat, fasilitas, dan telah membuka jalan bagi penulis untuk lebih banyak berperan dalam berbagai kegiatan yang membawa nama almamater Fisika IPB sejak awal penulis masuk ke dalam keluarga Fisika IPB.
3. Bapak dan Mama dan adik tersayang yang selalu mendoakan penulis dan atas semua fasilitas yang diberikan semenjak penulis terlahir ke dunia ini dan selalu berharap agar penulis menjadi manusia yang terbaik di mata Allah *Subhanahuwata'ala*.
4. Kartika Andansari (*my dear*) yang selalu ada di kala penulis senang maupun sedih, yang selalu mendoakan dan memotivasi penulis hingga penulis dapat berbuat maksimal atas semua hal yang penulis impikan hingga saat ini (*for our best future, dear*).
5. Moh. Rosyid Mahmudi, sahabat karib penulis dan juga teman satu topik penelitian yang telah memberikan teladan kepada penulis berupa kerja keras tanpa mengenal putus asa.
6. Para penghuni laboratorium Fisika Teori dan Komputasi IPB: Rudiyanto, Andrial S, Fabian R, Izzatu Y, kak Teguh P N (Fis'39), kak Mardanih (Fis'40), Pak Mamat (Fis'32), dll.
7. Teman-teman satu Departemen Fisika angkatan 43 : Ridwan, Nady, (Alm.) Bobby Novian, Mila, Afni, Yulis, Dina, Mufti, Galih, Husein, Sastri, Rendra, Yuli dan lain-lain yang tidak dapat penulis sebutkan satu per satu. *You're all my best friends ever*.
8. Teman-teman fisika angkatan 40, 41, 42, 44, 45, teman-teman satu kost di pondok *Pioneer*, teman-teman alumni SMUN 47, Iqbal (Bio'43), Kak Yohan (KPM'39), dan lain-lain yang telah banyak berkontribusi dengan penulis selama masa studi serta penelitian di departemen Fisika IPB.
9. Semua Dosen, Staf, dan Karyawan departemen Fisika IPB yang telah mengajarkan banyak hal kepada penulis hingga pencapaian penulis seperti sekarang ini.

Penulis menyadari bahwa penelitian ini masih jauh dari sempurna sehingga segala saran dan kritik yang membangun bagi penulis dan pada penelitian ini sangat penulis harapkan. Semoga kita selalu dalam naungan rahmat dan hidayah Allah *subhanahuwata'ala*. *Amiin*.

*Wassalamu'alaikum warahmatullahi wabarakatuh*

Bogor, Januari 2010

Penulis

# DAFTAR ISI

# DAFTAR GAMBAR

# DAFTAR LAMPIRAN

# PENDAHULUAN

### Latar Belakang

Beberapa dekade belakangan ini, perhatian terbesar bidang fisika-optik bertujuan untuk mengendalikan sifat-sifat optis material dengan aplikasi yang sangat menjanjikan di masa depan. Kemampuan dalam mengendalikan dan melokalisasi penjalaran cahaya menghasilkan teknologi serat optik, laser, dan komunikasi berkecepatan tinggi yang telah mengubah tren masa depan teknologi informasi.

Kristal fotonik adalah material dielektrik yang memiliki indeks bias atau permitivitas berbeda secara periodik, sehingga dapat mencegah perambatan cahaya dengan frekuensi dan arah tertentu. Rentang daerah frekuensi tersebut dinamakan *photonic bandgap* (PBG). Dasar teoretis mengenai kristal fotonik dikembangkan pertama kali oleh E. Yablonovic dan S. John pada tahun 1987. Sebagai hasilnya dijelaskan bahwa banyak fenomena yang terjadi pada semikonduktor terjadi pula pada bahan kristal fotonik. Jika interval frekuensi dan sudut datang berbeda-beda, maka cahaya akan dipantulkan seluruhnya oleh kristal fotonik sehingga diketahui sifat-sifat fotonik *bandgap* secara lengkap.

Suatu sifat yang penting pada kristal fotonik dapat diperoleh jika terdapat cacat (defek) pada struktur kristal fotonik. Defek tersebut menimbulkan keadaan terlokalisasi di sekitar *bandgap* sehingga hanya akan terjadi transmitansi pada satu atau beberapa selang frekuensi tertentu yang disebut sebagai *bandpass*. Variasi pada struktur defek menghasilkan modifikasi pada frekuensi *bandpass* sedangkan jika struktur kristal fotonik dimodifikasi dengan mengambil satu baris silinder (*rod*) dalam kristal fotonik maka didapatkan sebuah pandu gelombang (*waveguide*). Dengan variasi struktur cacat di sekitar pandu gelombang maka akan terjadi pengalihan sebagian atau keseluruhan (kopling) dari medan elektromagnetik (TM) yang dirambatkan pada kristal fotonik tersebut.

Pada penelitian ini variasi cacat dibatasi hanya pada model dua cacat simetris di sekitar pandu gelombang sehingga dapat dianalisis karakter dari masing-masing model cacat tersebut.

### Tujuan

Tujuan penelitian ini adalah untuk menganalisis karakteristik kopling pada kristal fotonik 2D dengan cacat simetris di sekitar sebuah kanal pandu gelombang dan menganalisis frekuensi-frekuensi efektif dan energi medan listrik maksimum dalam kopling tersebut.

# TINJAUAN PUSTAKA

### Kristal Fotonik 2D Sederhana

Kristal fotonik 2D tersusun periodik pada dua sumbu aksisnya, dan homogen sepanjang sumbu ketiga. Kristal fotonik 2D terdiri atas susunan kolom-kolom dielektrik seperti yang ditunjukkan pada **Gambar 2.** Pada kristal foronik terdapat nilai *bandgap* pada bidang x-y. Di dalam *gap* ini, tidak ada transmitasnsi yang terjadi, dan cahaya yang menumbuknya akan dipantulkan seluruhnya. Kristal fotonik 2D dapat memantulkan cahaya yang datang dari arah manapun pada bidang sehingga tidak ada cahaya yang dapat ditransmisikan di dalamnya.

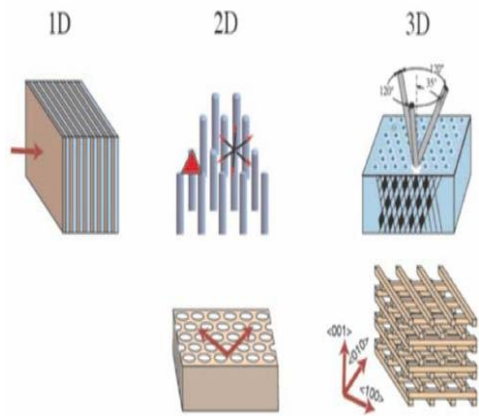

**Gambar 1** Struktur Kristal fotonik 1D, 2D, dan 3D [7].

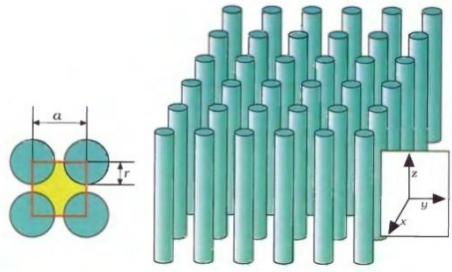

**Gambar 2** Kristal fotonik 2D yang terdiri atas kolom-kolom dielektrik dengan jari-jari $r$ dan konstanta dielektrik $\varepsilon_r$. Material ini homogen pada sumbu $z$ (digambarkan seperti silinder yang panjang), dan periodik pada bidang $x$-$y$ dengan konstanta jarak pada pusat silinder sebesar $a$. [4]

Jika $k_z = 0$, maka cahaya merambat pada bidang $x$-$y$. Gelombang EM yang datang pada kristal fotonik 2D dapat mengklasifikasikan modus perambatannya dengan membagi menjadi dua polarisasi yang berbeda. Modus *transverse electric* (TE) memiliki arah medan magnet $H$ yang tegak lurus terhadap bidang, $H = H(\rho)\hat{z}$ dan medan listrik E yang sejajar dengan bidang, $E(\rho).\hat{z} = 0$. Modus *transverse magnetic* (TM) bersifat sebaliknya, yaitu $E = E(\rho)\hat{z}$ dan $H(\rho).\hat{z} = 0$ [4].

**Persamaan Gelombang Datar pada Kristal Fotonik Isotropik**

Secara klasik gelombang Elektromagnetik (EM) tersusun atas dua vektor medan yaitu medan magnet H dan medan listrik E. Kedua vektor medan tersebut berosilasi saling tegak lurus dan merambat tanpa membutuhkan medium secara transversal (arah osilasi ⊥ arah rambat). Perambatan kedua vektor medan ini dijelaskan dengan baik melalui persamaan-persamaan Maxwell. Dengan menyertakan pengaruh medium terhadap kedua medan tersebut, maka persamaan-persamaan Maxwell menjadi berbentuk :

$$\vec{\nabla} \times \vec{E} = -\frac{\partial}{\partial t}\vec{B} \qquad (2.1)$$

$$\vec{\nabla} \times \vec{H} = \frac{\partial}{\partial t}D + J \qquad (2.2)$$

$$\vec{\nabla} \cdot \vec{B} = 0 \qquad (2.3)$$

$$\vec{\nabla} \cdot \vec{D} = \rho \qquad (2.4)$$

dengan $J$ merupakan rapat arus listrik (Ampere per meter persegi) dan $\rho$ merupakan rapat muatan listrik (Coulomb per meter kubik). Serta vektor yang menyatakan interaksi dengan medium, yaitu vektor perpindahan listrik $D$ dan induksi magnetik $B$. Keempat persamaan ini merupakan hukum dasar kelistrikan dan kemagnetan dalam bentuk differensialnya. Persamaan (2.1) perupakan persamaan hukum faraday tentang induksi magnetik, yang menggambarkan pembentukan medan listrik induksi rotasional akibat adanya perubahan fluks magnetik terhadap waktu. Persamaan (2.2) merupakan bentuk differensial dari hukum Ampere yang diperumum dan menggambarkan timbulnya medan magnet induksi rotasional akibat adanya muatan listrik yang mengalir pada suatu penghantar. Persamaan (2.3) menyatakan bahwa divergensi dari fluks magnet bernilai nol. Dengan kata lain, hingga saat ini tidak terdapat atau setidaknya belum ditemukan adanya monopol magnet di alam semesta. Persamaan (2.4) merupakan bentuk differensial dari hukum Coulomb yang menyatakan hubungan antara distribusi medan listrik yang ditimbulkan oleh suatu distribusi muatan.

Baik rapat muatan, $\rho$ maupun rapat arus, $J$ dapat dinyatakan sebagai sumber radiasi EM. Pada penelitian ini radiasi EM dirambatkan melalui medium yang jauh dari sumber sehingga $\rho$ dan $J$ dapat dianggap nol.

Keempat persamaan Maxwell tersebut membentuk suatu persamaan differensial parsial yang terkopel melalui keempat kuantitas medan $E, H, D, B$. oleh karena itu, dibutuhkan persamaan konstitutif untuk menentukan vektor medan dari sekumpulan distribusi arus dan muatan,

$$\vec{D} = \varepsilon\vec{E} = \varepsilon_0\vec{E} + \vec{P} \qquad (2.5)$$
$$\vec{B} = \mu\vec{H} = \mu_0\vec{H} + \vec{M} \qquad (2.6)$$

dengan $\varepsilon$ dan $\mu$ merupakan tensor dielektrik (tensor permitivitas) dan tensor permeabilitas. $P$ dan $M$ adalah polarisasi listrik dan magnetik, serta $\varepsilon_0$ merupakan permitivitas dalam vakum dan $\mu_0$ adalah permeabilitas dalam vakum. Kuantitas $\varepsilon$ dan $\mu$ bergantung pada $E$ dan $H$ jika medan tersebut cukup kuat. Kebergantungan $\varepsilon$ dan $\mu$ dalam medan yang cukup kuat bersifat nonlinier akan tetapi pada penelitian ini medan yang digunakan tidak terlalu kuat sehingga kebergantungan kedua kuantitas tadi terhadap *E* dan *H* dapat diabaikan.

Pada kristal fotonik 2 dimensi yang bersifat homogen (kerapatan sama), isotropik (sama ke semua arah), dan jauh dari sumber muatan listrik dan magnet (nonmagnetik; *M* = 0) maka berlaku persamaan konstitutif . Subtitusi ke dalam persamaan Maxwell menjadi,

$$\vec{\nabla}\times\vec{E}+\frac{\partial}{\partial t}\mu\vec{H}=0 \qquad (2.7)$$

$$\vec{\nabla}\times\vec{H}-\frac{\partial}{\partial t}\varepsilon\vec{E}=0 \qquad (2.8)$$

$$\vec{\nabla}\cdot\mu\vec{H}=0 \qquad (2.9)$$

$$\vec{\nabla}\cdot\varepsilon\vec{E}=0 \qquad (2.10)$$

Persamaan gelombang harmonik dapat dituliskan dalam bentuk eksponensial kompleks: $\vec{E}(\vec{r},t)=\vec{E}(\vec{r})e^{-i\omega t}$ dan $\vec{H}(\vec{r},t)=\vec{H}(\vec{r})e^{-i\omega t}$. Subtitusikan ke dalam persamaan (2.7) maka,

$$\vec{\nabla}\times\vec{E}(\vec{r})e^{-i\omega t}-i\mu\omega\vec{H}(\vec{r})e^{-i\omega t}=0 \quad (2.11)$$

dengan mengambil suku realnya menjadi,

$$Re[\vec{\nabla}\times\vec{E}(\vec{r})-i\mu\omega\vec{H}(\vec{r})]e^{-i\omega t}=0 \quad (2.12)$$

Karena suku $e^{-i\omega t}$ tidak bernilai nol, maka suku di dalam kurung harus bernilai nol

$$\vec{\nabla}\times\vec{E}(\vec{r})-i\mu\omega\vec{H}(\vec{r})=0 \qquad (2.13)$$

Dengan cara yang sama maka persamaan (2.8) menjadi berbentuk,

$$\vec{\nabla}\times\vec{H}(\vec{r})+i\varepsilon\omega\vec{E}(\vec{r})=0 \qquad (2.14)$$

Kemudian jika persamaan (2.13) dikenakan operasi *Curl*, maka akan didapatkan persamaan,

$$\vec{\nabla}\times\vec{\nabla}\times\vec{E}(\vec{r})=\vec{\nabla}\times i\mu\omega\vec{H}(\vec{r})$$

$$\vec{\nabla}\times\vec{\nabla}\times\vec{E}(\vec{r})=i\mu\omega[\vec{\nabla}\times\vec{H}(\vec{r})]$$

$$\vec{\nabla}\times\vec{\nabla}\times\vec{E}(\vec{r})-\varepsilon\mu\omega^2\vec{E}(\vec{r})=0 \quad (2.15)$$

Dengan menggunakan hubungan $k=\omega\sqrt{\varepsilon\mu}$, maka persamaan (2.15) menjadi,

$$\vec{\nabla}\times\vec{\nabla}\times\vec{E}(\vec{r})-k^2\vec{E}(\vec{r})=0 \qquad (2.16)$$

Sekarang ditinjau bila kristal fotonik terdiri atas dua medium. Medium pertama berbentuk silider yang memiliki fungsi dielektrik *ε(r)* berada pada medium dasar (*background*) dengan permitivitas *ε_B*. Jika medium dasar tidak vakum dan nonmagnetik, total medan listrik dari gelombang datang $E^0(\vec{r})$ yang menjalar pada medium dasar merupakan solusi dari persamaan vektor gelombang,

$$\vec{\nabla}\times\vec{\nabla}\times\vec{E}(\vec{r})-k_0^2\varepsilon(\vec{r})\vec{E}(\vec{r})=0 \quad (2.17)$$

dengan $k=k_0\sqrt{\varepsilon(\vec{r})}$ dan $k_0$ adalah bilangan gelombang dalam vakum. Selisih konstanta dielektrik antara dua medium adalah $\Delta\varepsilon(\vec{r})=\varepsilon(\vec{r})-\varepsilon_B$. Dengan menggunakan selisih konstanta dielektrik tersebut, maka persamaan gelombang vektor homogen diubah menjadi bentuk yang tak homogen [2],

$$\vec{\nabla}\times\vec{\nabla}\times\vec{E}(\vec{r})-k_0^2\varepsilon_B\vec{E}(\vec{r})=k_0^2\Delta\varepsilon(\vec{r})\vec{E}(\vec{r})$$
(2.18)

Persamaan (2.18) akan digunakan untuk menurunkan tensor Green pada sistem dua dimensi.

**Fungsi Green Dyadic pada Medium Homogen**

Fungsi Green dyadic terdiri atas dyad yang menghubungkan medan vektor dengan sumber vektor arus. Pada medium isotropik, homogen, dan jauh dari sumber muatan maka Medan Listrik $\vec{E}(\vec{r})$ dipenuhi menurut persamaan,

$$\vec{\nabla}\times\vec{\nabla}\times\vec{E}(\vec{r})-k^2\vec{E}(\vec{r})=0 \quad (2.19)$$

Dengan menggunakan identitas vektor,

$$\vec{\nabla}\times\vec{\nabla}\times\vec{E}=\vec{\nabla}(\vec{\nabla}\cdot\vec{E})-\vec{\nabla}^2\vec{E}$$

Kemudian persamaan (2.10) menjadi,

$$\vec{\nabla}\cdot\varepsilon\vec{E}=\varepsilon(\nabla\cdot\vec{E})+(\vec{E}\cdot\nabla\varepsilon)=0$$

Dalam medium homogen dan isotropik, maka gradien dari *ε* bisa dianggap nol sehingga suku kedua bernilai nol, sehingga, $\varepsilon(\nabla\cdot\vec{E})=0$ atau $\vec{\nabla}\cdot\vec{E}=0$. Persamaan (2.19) dengan demikian menjadi,

$$\vec{\nabla}^2\vec{E}(\vec{r})+k^2\vec{E}(\vec{r})=0 \qquad (2.20)$$

Persamaan tersebut dikenal sebagai persamaan Helmholtz yang merupakan persamaan gelombang EM standar yang menggambarkan perambatan gelombang EM dalam kristal fotonik isotropik homogen dan jauh dari sumber muatan (bebas sumber).

Fungsi Green pada gelombang merupakan solusi persamaan gelombang oleh sumber titik. Jika solusi persamaan gelombang pada satu sumber titik diketahui, maka solusi dari beberapa sumber titik didapatkan melalui superposisi linier dari masing-masing sumber titik tersebut. Untuk mendapatkan solusi skalar persamaan gelombang dari satu titik sumber maka digunakan penurunan berikut ini. Dengan menggunakan persamaan gelombang,

$$(\vec{\nabla}^2 + k^2)\vec{E}(\vec{r}) = s(\vec{r}) \qquad (2.21)$$

maka dapat ditentukan fungsi Greennya yang merupakan solusi dari,

$$(\vec{\nabla}^2 + k^2)g(\vec{r},\vec{r}') = -\delta(\vec{r} - \vec{r}') \qquad (2.22)$$

Untuk menyelesaikan persamaan gelombang skalar, maka persamaan (2.22) berkaitan dengan $\vec{E}(\vec{r})$ melalui hubungan,

$$\vec{E}(\vec{r}) = \int_V d\mathrm{r}' g(\vec{r},\vec{r}') s(\vec{r}') \qquad (2.23)$$

Pada sumber titik, akan lebih memudahkan jika digunakan sistem koordinat bola dengan titik asal $\vec{r}'$ sehingga persamaan (2.23) berbentuk,

$$(\vec{\nabla}^2 + k^2)g(\vec{r}) = -\delta(\vec{r}) \qquad (2.24)$$

dengan koordinat bola yang mengambil bentuk,

$$\left\{\left[\frac{1}{r}\frac{\partial^2}{\partial r^2}r + \frac{1}{r^2 \sin\theta}\frac{\partial}{\partial\theta}\left(\sin\theta\frac{\partial}{\partial\theta}\right) + \frac{1}{r^2 \sin^2\theta}\frac{\partial}{\partial\varphi^2}\right] + k^2\right\}g(\vec{r}) = 0$$

Karena persamaan gelombang isotropik homogenik, maka didapatkan bentuk yang lebih sederhana,

$$\frac{\partial^2}{\partial r^2}\vec{r}\,g(\vec{r}) + k^2\vec{r}\,g(\vec{r}) = 0 \qquad (2.25)$$

Solusi umum dari persamaan (2.25) di atas adalah,

$$g(\vec{r}) = \frac{a}{\vec{r}}e^{ik\vec{r}} + \frac{b}{\vec{r}}e^{-ik\vec{r}} \qquad (2.26)$$

Karena keberadaan sumber dianggap tak hingga, maka hanya suku pertama yang digunakan,

$$g(\vec{r}) = \frac{a}{\vec{r}}e^{ik\vec{r}} \qquad (2.27)$$

Konstanta $a$ ditentukan dengan mensubtitusikan persamaan (2.27) pada persamaan (2.23) kemudian mengintegralkannya dengan menganggap $\delta(\mathrm{r}) = 1$ pada titik asalnya ($\vec{r}' = \vec{r}$),

$$\int_{\Delta V} dV (\nabla \cdot \nabla \frac{a}{\vec{r}}e^{ik\vec{r}} + k^2 \frac{a}{\vec{r}}e^{ik\vec{r}}) = -1$$

$$\int_{\Delta V} dV (\nabla \cdot \nabla \frac{a}{\vec{r}}e^{ik\vec{r}}) = \oint_{\Delta S} \nabla \frac{a}{\vec{r}}e^{ik\vec{r}} \cdot dS = -1$$

$$\lim_{\vec{r}\to 0} 4\pi r^2 \frac{\partial}{\partial r}\frac{a}{\vec{r}}e^{ik\vec{r}} = -1$$

$$a = \frac{1}{4\pi} \qquad (2.28)$$

sehingga fungsi Green dyadic skalar menjadi [2],

$$g(\vec{r}) = \frac{1}{4\pi|\vec{r}-\vec{r}'|}e^{ik|\vec{r}-\vec{r}'|} \qquad (2.29)$$

**Tensor Green 3D**

Diketahui tensor Green untuk medium 3D homogen berbentuk [8],

$$\vec{G}(\vec{r},\vec{r}') = \left[\vec{I} + \frac{\vec{\nabla}\vec{\nabla}}{k_B^2}\right]g(\vec{r},\vec{r}') \qquad (2.30)$$

Penurunan detail mengenai tensor Green 3D diberikan pada **Lampiran 1**. Fungsi Green dyadic yang ditunjukkan pada persamaan (2.29) dapat sedikit dimodifikasi menjadi berbentuk,

$$g(\vec{r}) = \frac{1}{4\pi R}e^{ikR} \qquad (2.31)$$

dengan $R = |\vec{r} - \vec{r}'|$ merupakan jarak mutlak antara sumber *(r')* dan pengamat *(r)*. konstanta $k_B$ merupakan bilangan gelombang pada medium *background* yang memiliki hubungan dengan permitivitas *background* $\varepsilon_B$ berbentuk,

$$k_B = \sqrt{\frac{\omega^2}{c^2}\varepsilon_B} \qquad (2.32)$$

Sistem tensor Green 3D homogen dapat dituliskan berbentuk,

$$\vec{G}^B(\vec{r},\vec{r}') = \begin{pmatrix} G_{xx} & G_{xy} & G_{xz} \\ G_{yx} & G_{yy} & G_{yz} \\ G_{zx} & G_{zy} & G_{zz} \end{pmatrix} \quad (2.34)$$

**Tensor Green pada Medium Homogen 2D**

Gambar 3 menunjukkan silinder yang dapat menghamburkan gelombang EM. Silinder tersebut memiliki sifat geometri yang homogen sepanjang sumbu-*z*. Jika arah gelombang EM sejajar dengan bidang penjalaran maka disebut sebagai polarisasi-*p*, sedangkan arah penjalaran gelombang yang tegak lurus terhadap bidang disebut sebagai polarisasi-*s*. Pada kasus khusus, jika sudut datang gelombang $\theta_i = 90^0$, maka komponen *transverse listrik* (TE) menunjukkan polarisasi-*p*, sedangkan komponen *transverse magnetic* (TM) menunjukkan polarisasi-*s*.

Melalui ilustrasi tersebut bila sudut datang gelombang $\theta_i = 90^0$, maka penjalaran gelombang EM pada silinder tersebut merupakan perambatan modus TM, dengan menganggap penjalaran TE hanya pada arah sumbu-*z* dan $k_B = k_\rho$.

Tensor Green 2D adalah distribusi medan pada bidang pengamatan dengan asumsi *z* = konstan sehingga fungsi Green skalar 2D dapat dibentuk dari bentuk 3D nya dengan menjumlahkan seluruh titik sumber yang terletak pada garis sepanjang sumbu-*z* atau hasil integrasi dari sebuah sumber garis,

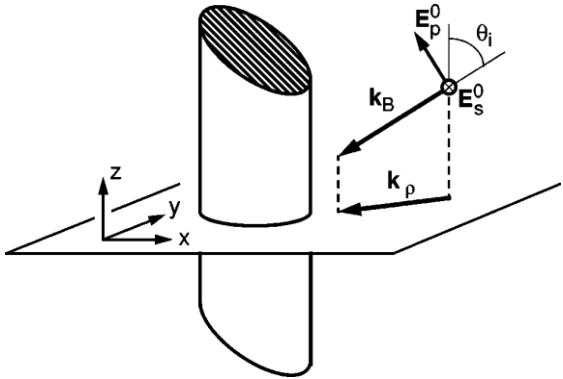

**Gambar 3** Sistem 2D [8].

$$g^B_{2D}(\vec{r},\vec{r}') = \int_{-\infty}^{\infty} dz' g^B_{3D}(\vec{r},\vec{r}') e^{ik_z z} \quad (2.35)$$

Dengan koordinat relatif $\rho^2 = x^2 + y^2$ asumsi bahwa sumber garis tersebut terletak pada $x' = y' = 0$. Sehingga,

$$g^B_{2D}(\vec{r},\vec{r}') = \int_{-\infty}^{\infty} dz' \frac{\exp[ik_B\sqrt{\rho^2+(z-z')^2}]}{4\pi\sqrt{\rho^2+(z-z')^2}} e^{ik_z z}$$

$$= \frac{i}{4} H_0(k_a, a) \exp[ik_z z] \quad (2.36)$$

Penurunan persamaan (2.36) secara rinci ditunjukkan dalam **Lampiran 2.** Komponen-komponen tensor Green 2D homogen didapatkan melalui penerapan pada persamaan dengan *z* = 0,

$$G^B(\vec{\rho},\vec{\rho}') = \begin{pmatrix} G^B_{xx} & G^B_{xy} & G^B_{xz} \\ G^B_{yx} & G^B_{yy} & G^B_{yz} \\ G^B_{zx} & G^B_{zy} & G^B_{zz} \end{pmatrix} \quad (2.37)$$

Dengan [8],

$$\vec{G}^B{}_{xx} = \frac{i}{4}\left(1 - \frac{k_\rho^2 \cos^2\phi}{k_B^2}\right) H_0(k_\rho,\rho) + \frac{i}{4}\frac{k_\rho^2 \cos 2\phi}{k_B^2 \rho} H_1(k_\rho,\rho)$$

$$\vec{G}^B{}_{xy} = \frac{i}{4} \frac{k_\rho^2 \sin 2\phi}{2k_B^2} H_2(k_\rho,\rho)$$

$$\vec{G}^B{}_{xz} = \frac{1}{4} \frac{k_z k_\rho \cos\phi}{k_B^2} H_1(k_\rho,\rho)$$

$$\vec{G}^B{}_{yy} = \frac{i}{4}\left(1 - \frac{k_\rho^2 \sin^2\phi}{k_B^2}\right) H_0(k_\rho,\rho) - \frac{i}{4}\frac{k_\rho^2 \cos 2\phi}{k_B^2 \rho} H_1(k_\rho,\rho)$$

$$\vec{G}^B{}_{yz} = \frac{1}{4} \frac{k_z k_\rho \sin\phi}{k_B^2} H_1(k_\rho,\rho)$$

$$\vec{G}^B{}_{zz} = \frac{i}{4}\left(1 - \frac{k_z^2}{k_B^2}\right) H_0(k_\rho,\rho) \quad (2.38)$$

Penurunan masing-masing fungsi sebagai komponen tensor Green ditunjukkan pada **Lampiran 2.** Jika medan elektromagnetik menjalar pada bidang x-y dan dengan menggunakan sifat simetri tensor, $G^B(\vec{r},\vec{r}') = G^B(\vec{r}',\vec{r})$, maka tensor Greennya menjadi,

$$G^B(\vec{\rho},\vec{\rho}') = \begin{pmatrix} G^B_{xx} & G^B_{xy} & 0 \\ G^B_{xy} & G^B_{yy} & 0 \\ 0 & 0 & G^B_{zz} \end{pmatrix}$$

(2.39)

Pada kasus polarisasi TM, tensor Green $G^B(\vec{\rho},\vec{\rho}')$ disederhanakan menjadi fungsi Green skalar dengan $z = 0$,

$$G^B(\vec{\rho},\vec{\rho}') = G^B_{zz}$$

(2.40)

Karena gelombang datang pada bidang $x - y$, kemudian $k_\rho = k_B$ dan $k_z = 0$, sehingga,

$$\vec{G}^B_{zz} = \frac{i}{4} H_0(k_\rho,\rho)$$ (2.41)

**Total Medan Listrik $\vec{E}(\vec{r})$**

Persamaan (2.18) menunjukkan persamaan vektor gelombang yang ditulis sebagai persamaan inhomogen,

$$\vec{\nabla}\times\vec{\nabla}\times\vec{E}(\vec{r}) - k_0^2 \varepsilon_B \vec{E}(\vec{r}) = k_0^2 \Delta\varepsilon(\vec{r})\vec{E}(\vec{r})$$

Total medan dapat dihitung dengan bantuan penurunan persamaan vektor fungsi Green,

$$\nabla\times\nabla\times\overline{G}(r,r') - k_0^2 \overline{G}(r,r') = \overline{I}\delta(r-r')$$ (2.42)

Karena solusi dari total medan $\vec{E}(\vec{r})$ juga mengandung solusi dari medan sumber awal $\vec{E}^0$ atau komponen homogen dan komponen inhomogen, dapat dibuktikan bahwa dengan melakukan subtitusi persamaan (2.42) ke persamaan (2.18) maka total medan $\vec{E}(\vec{r})$ akan berbentuk [2],

$$\vec{E}(\vec{r}) = \vec{E}^0(\vec{r}) + \int_V dr' G^B(\vec{r},\vec{r}')\cdot k_0^2 \Delta\varepsilon(\vec{r}')\vec{E}(\vec{r}')$$

(2.43)

dengan *V* adalah volume hamburan total.

Diketahui nilai skalar fungsi Green pada persamaan (2.18) bernilai singular jika lokasi pengamat sama dengan letak sumber $(\vec{r} = \vec{r}')$, sehingga tensor Green juga akan bernilai singular. Untuk memecahkan kesulitan pada kasus $\vec{r} = \vec{r}'$, sifat singularitas dari tensor Green harus dipecahkan menjadi bentuk yang lebih sederhana. Modifikasi dari persamaan (2.43) menjadi bentuk persamaan lipmann-Schwinger,

$$\vec{E}(\vec{r}) = \vec{E}^0(\vec{r})$$
$$+ \lim_{\delta V \to 0} \int_{V-\delta V} dr' G^B(\vec{r},\vec{r}')\cdot k_0^2 \Delta\varepsilon(\vec{r}')\vec{E}(\vec{r}') - \vec{L}\cdot\frac{\Delta\varepsilon(\vec{r})}{\varepsilon_B}\vec{E}(\vec{r})$$

(2.44)

Integral dalam persamaan tersebut mengeksklusi singularitas dengan konsekuensinya adalah penambahan komponen dyadic $\vec{L}$ yang bergantung dengan batas volume eksklusi $\delta V$. Komponen dari $\vec{L}_i$ yang menghubungkan sumber dyadic pada batang silinder 2D dalam kasus ini [8],

$$\vec{L}_i = \begin{pmatrix} \frac{1}{2} & 0 & 0 \\ 0 & \frac{1}{2} & 0 \\ 0 & 0 & 0 \end{pmatrix}$$

(2.45)

Pada kasus modus TM maka nilai $\vec{L}$ adalah nol $(\vec{L}_{zz} = 0)$.

## METODOLOGI

**Tempat dan Waktu Penelitian**

Penelitian ini dilakukan di Laboratorium Fisika Teori dan Komputasi, Departemen Fisika, FMIPA, IPB.

Penelitian ini dilakukan selama 6 bulan, dimulai dari bulan Agustus 2009 sampai Januari 2010 dan meliputi studi literatur yang terkait, pembuatan program simulasi kristal fotonik 2D, analisis hasil simulasi program, dan penulisan laporan akhir. Penelitian ini penulis lakukan disela-sela waktu kuliah yaitu kuliah semester 7.

**Peralatan**

Peralatan yang digunakan dalam penelitian ini adalah sebuah laptop berprosesor Intel Dual Core (2.0 GHz), 2 GB RAM dan komputer PC berprosesor Intel Quad Core, 2 GB RAM. Software yang digunakan dalam penelitian ini adalah Ms. Office 2007 dan MATLAB R2008a / R2008b. Sebagai mendukung penelitian ini penulis menggunakan berbagai sumber literatur, yaitu berupa buku teks dan jurnal-jurnal ilmiah.

**Metode Penelitian**

Metode yang digunakan pada penelitian ini adalah penghitungan medan listrik total menggunakan tensor Green. Perhitungan dilakukan secara numerik dengan menggunakan persamaan Lippman-Schwinger yang terdiskretisasi.

**Pembuatan *Mesh* untuk Satu Silinder**

Pada penelitian ini dibuat susunan silinder yang terdiri atas sebuah silinder tunggal yang disusun periodik sebagai representasi dari sebuah silinder tunggal pada kristal fotonik 2D.

Sebuah *mesh* didefinisikan sebagai suatu kotak kecil yang mengandung nilai permitivitas tertentu sehingga akan dibentuk susunan mesh yang berbentuk sebuah silinder tunggal dengan permitivitas bahan tertentu pada suatu medium dengan permitivitas sama dengan udara. Bentuk susunan mesh untuk satu silinder ditunjukkan pada gambar berikut:

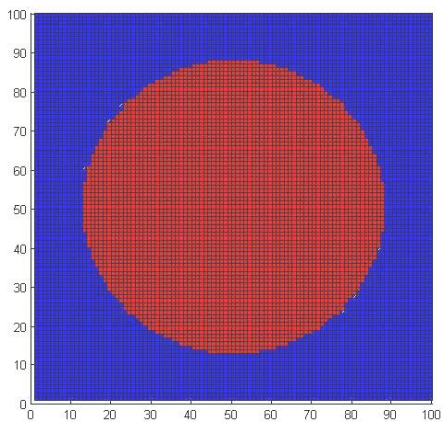

**Gambar 4** Struktur satu silinder Kristal fotonik 2D

Pada penelitian ini digunakan silinder (*Rod*) dengan permitivitas bahan $\varepsilon_r = 11{,}56$ dengan medium udara $\varepsilon = 1$. Silinder yang digunakan memiliki jari-jari sebesar 0,25 µm.

**Pembentukkan Gelombang Datar (*Planewave*)**

Penelitian ini menggunakan gelombang EM dengan modus *transverse magnetic* (TM) yang akan dirambatkan pada kristal fotonik 2D.

Gelombang EM yang digunakan dalam penelitian ini adalah gelombang datar yang bebas waktu,

$$E_0 = Ae^{-ikx} \qquad (3.1)$$

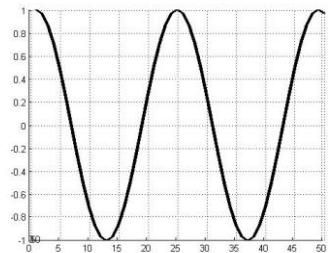

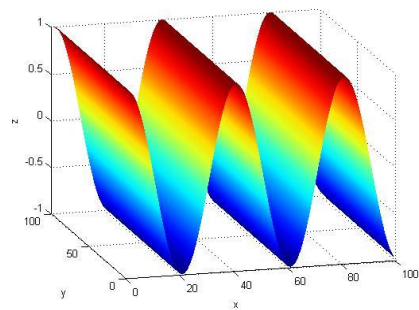

**Gambar 5** gelombang datar EM modus *transverse magnetic* (TM)

Gelombang EM yang digunakan dalam penelitian ini memiliki frekuensi bervariasi dengan interval dari $\frac{\omega a}{2\pi c} = 0.36$ sampai $\frac{\omega a}{2\pi c} = 0{,}39$. Frekuensi tersebut berada dalam interval *bandgap* berdasarkan data literatur yang digunakan.

**Diskretisasi Persamaan Lippman-Schwinger**

Persamaan lippman-Schwinger dapat diselesaikan secara numerik dengan memanfaatkan bentuk diskritnya. Dalam melakukan pendekatan tersebut maka dibuat sejumlah N kotak-kotak kecil. Masing-masing kotak memiliki volume yang konstan $V_i$. Dapat dituliskan bentuk diskrit dari persamaan tersebut adalah,

$$\vec{E}_i = \vec{E}_i^0 + \sum_{j=1, j\neq 1}^{N} \vec{G}_{ij}^B \cdot k_0^2 \Delta\varepsilon_j \vec{E}_j V_j + \vec{M} \cdot k_0^2 \Delta\varepsilon_i \vec{E}_i - \vec{L} \cdot \frac{\Delta\varepsilon_i}{\varepsilon_B} \vec{E}_i \quad (3.2)$$

dengan $\vec{E}_i = \vec{E}(r_i)$ adalah medan terdiskretisasi dan $\Delta\varepsilon_i = \Delta\varepsilon_i(\vec{r}_i)$ adalah selisih konstanta dielektrik terdiskretisasi. Diskretisasi ini dapat bervariasi untuk mendapatkan tingkat akurasi yang diinginkan, dengan ukuran kotak yang lebih kecil sedangkan selisih konstanta dielektrik $\Delta\varepsilon(\vec{r})$ yang besar.

Komponen $\vec{M}$ muncul sebagai akibat peralihan dari bentuk integrasi kedalam bentuk penjumlahan. Untuk mendapatkan pemahaman yang lebih baik mengenai $\vec{M}$ lihat Gambar 6. Keberadaan komponen ini kurang signifikan dibandingkan dengan komponen lainnya sehingga sering diabaikan. Walaupun demikian, komponen tersebut mempunyai peranan yang sangat penting dalam meningkatkan akurasi numerik.

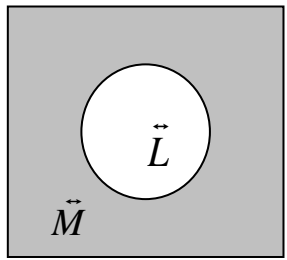

**Gambar 6**. Definisi komponen $\vec{M}$ yaitu daerah yang berwarna kelabu dari kotak dengan volume $V_i$ dikurangi dengan daerah komponen dyadic $\vec{L}$

Besar nilai komponen $M_i$ dapat dihitung dengan [8],

$$\vec{M}_i = \lim_{\delta V \to 0} \int_{V_i - \delta V} d\vec{r}' \vec{G}^B(\vec{r}_i, \vec{r}) \quad (3.3)$$

atau dalam bentuk yang eksplisit menjadi:

$$\vec{M}_i = \begin{pmatrix} \frac{i\pi}{4}\alpha\gamma & 0 & 0 \\ 0 & \frac{i\pi}{4}\alpha\gamma & 0 \\ 0 & 0 & \frac{i\pi}{4}\beta\gamma \end{pmatrix} \quad (3.4)$$

dengan $\alpha = \left(2 - \frac{k_\rho^2}{k_B^2}\right)$, $\beta = \left(1 - \frac{k_z^2}{k_B^2}\right)$, dan

$\gamma = \left(\frac{R_i^{eff}}{k_\rho} H_1(k_\rho, R_i^{eff}) + \frac{2i}{\pi k_\rho^2}\right)$

pada kasus gelombang TM, maka komponen yang digunakan adalah $M_{zz} = \frac{i\pi}{4}\beta\gamma$ [8].

**Penghitungan Total Medan TM dengan Persamaan Lippman-Schwinger terdiskretisasi**

Persamaan Lippman-Swinger memiliki bentuk terdiskretisasi berupa,

$$\vec{E}_i = \vec{E}_i^0 + \sum_{j=1, j\neq 1}^{N} \vec{G}_{ij}^B \cdot k_0^2 \Delta\varepsilon_j \vec{E}_j V_j + \vec{M} \cdot k_0^2 \Delta\varepsilon_i \vec{E}_i - \vec{L} \cdot \frac{\Delta\varepsilon_i}{\varepsilon_B} \vec{E}_i$$

Bentuk persamaan tersebut digunakan untuk menghitung besarnya nilai total medan $E$ pada kristal fotonik 2D.

Persamaan Lippman-Schwinger menunjukkan bahwa nilai $\vec{E}_i$ dan $\vec{E}_j$ sebenarnya memiliki bentuk yang sama dan dihubungkan melalui persamaan,

$$\vec{E}_i = \delta_{ij} \vec{E}_i \quad (3.5)$$

Sehingga persamaan Lippman-Schwinger berubah menjadi,

$$\vec{E}_i = \vec{E}_i^0 + \sum_{j=1, j\neq 1}^{N} \vec{G}_{ij}^B \cdot k_0^2 \Delta\varepsilon_j \vec{E}_j V_j$$
$$+ \vec{M} \cdot k_0^2 \Delta\varepsilon_i \vec{E}_i - \vec{L} \cdot \frac{\Delta\varepsilon_i}{\varepsilon_B} \vec{E}_i \quad (3.6)$$

$$\delta_{ij} \cdot \vec{E}_j = \vec{E}_i^0 + \sum_{j=1, j\neq 1}^{N} \vec{G}_{ij}^B \cdot k_0^2 \Delta\varepsilon_j \vec{E}_j V_j$$
$$+ \delta_{ij} \cdot (\vec{M} \cdot k_0^2 \Delta\varepsilon_i \vec{E}_j - \vec{L} \cdot \frac{\Delta\varepsilon_i}{\varepsilon_B} \vec{E}_j) \quad (3.7)$$

$$\delta_{ij} \cdot \vec{E}_j = \vec{E}_i^0 + (\sum_{j=1, j\neq 1}^{N} \vec{G}_{ij}^B \cdot k_0^2 \Delta\varepsilon_j V_j$$
$$+ \delta_{ij} \cdot (\vec{M} \cdot k_0^2 \Delta\varepsilon_i - \vec{L} \cdot \frac{\Delta\varepsilon_i}{\varepsilon_B})) \cdot \vec{E}_j$$

(3.8)

maka dapat didefinisikan sebuah matriks baru yang berbentuk,

$$T_{ij} = \sum_{j=1, j\neq 1}^{N} \vec{G}_{ij}^B \cdot k_0^2 \Delta\varepsilon_j V_j + \delta_{ij} \cdot (\vec{M} \cdot k_0^2 \Delta\varepsilon_i - \vec{L} \cdot \frac{\Delta\varepsilon_i}{\varepsilon_B})$$

(3.9)

sehingga bentuk persamaan Lippman-Schwinger menjadi,

$$\delta_{ij} \cdot \vec{E}_j = \vec{E}_i^0 + T_{ij} \cdot \vec{E}_j$$
$$\delta_{ij} \cdot \vec{E}_j - T_{ij} \cdot \vec{E}_j = \vec{E}_i^0$$
$$(\delta_{ij} - T_{ij}) \cdot \vec{E}_j = \vec{E}_i^0$$
$$\vec{E}_j = (\delta_{ij} - T_{ij})^{-1} \cdot \vec{E}_i^0$$
$$\vec{E}_j = (I_{ij} - T_{ij})^{-1} \cdot \vec{E}_i^0$$

(3.10)

dengan $\delta_{ij} = I_{ij}$ yang merupakan sebuah matriks identitas berordo $i \times j$. Maka medan total $\vec{E}_j$ dapat dihitung dengan persamaan tersebut.

### *Bandgap* pada Kristal Fotonik 2D

Kristal fotonik merupakan suatu susunan periodik dari material dengan permitivitas atau indeks bias tertentu yang dapat menghalangi perambatan gelombang EM yang datang dengan frekuensi dan arah tertentu yang dinamakan sebagai *photonic bandgap* (PBG).

Dalam penelitian ini gelombang datang EM TM pada kristal fotonik 2D yang disimulasikan berada pada frekuensi *bandgap* sehingga tak ada gelombang yang dapat menembus susunan silinder dalam kristal fotonik 2D dan arah perambatan gelombang EM dapat dikendalikan dengan memberikan kanal (*waveguide*) dan cacat (defek). Sumber data PBG yang digunakan ditunjukkan dalam gambar,

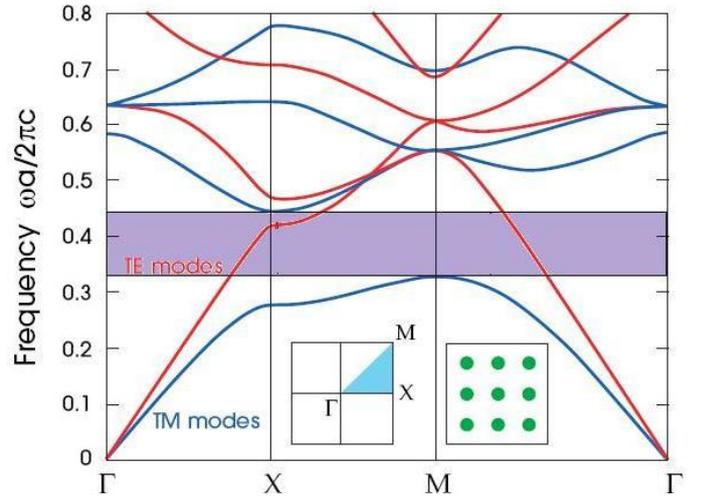

**Gambar 7**. *Photonic bandgap* pada Kristal fotonik 2D dengan parameter kisi $r = 0,2a$ [5].

| R (a) | Modus/Konstanta dielektrik silinder kisi segiempat | | | | | | Modus/Konstanta dielektrik silinder kisi segienam | | | | | |
|---|---|---|---|---|---|---|---|---|---|---|---|---|
| | modus TE | | | modus TM | | | modus TE | | | modus TM | | |
| | 8 | 11 | 14 | 8 | 11 | 14 | 8 | 11 | 14 | 8 | 11 | 14 |
| 0,2 | - | - | - | LB | LB | LB | - | - | - | LB | LB | LB |
| 0,3 | - | - | - | SD | SD | SD | - | SM | SM | SD | SD | SD |
| 0,4 | - | - | - | SM | SM | SM | - | SM | SM | SM | SM | SM |
| 0,45 | - | - | - | - | - | - | - | - | - | - | - | - |

SM : *band gap* < 5%  |  SD : 5% < *band gap* < 10%  |  LB : *band gap* > 10%  |  a : jarak antar kisi

**Gambar 8**. Hubungan *Photonic bandgap* dengan parameter kisi dan permitivitas bahan pada Kristal fotonik 2D [3].

Gambar 8 memperlihatkan bahwa *bandgap* berkurang dengan pertambahan jarak antar kisi pada kristal fotonik 2D. Pada penelitian ini jarak antar kisi (*a*) yang digunakan adalah sebesar $r = 0,2a$ dengan jari-jari dari tiap silinder berukuran 0,25 μm. Hal ini terlihat pada gambar di atas bahwa pada kisi segi empat modus TM pada penelitian ini, *bandgap* terbesar berada pada kisaran jarak antar kisi tersebut.

# HASIL DAN PEMBAHASAN

### Simulasi Hamburan Gelombang EM TM oleh Satu Silinder

Pada penelitian ini dibuat sebuah silinder yang berada dalam sebuah medium dengan permitivitas tertentu. Pembuatan silinder ini bertujuan untuk menyesuaikan program MATLAB yang dibuat dengan hasil simulasi dari literatur referensi.

Silinder (*rod*) yang digunakan pada hamburan ini memiliki permitivitas $\varepsilon_r = 4$ dan medium (*background*) yang digunakan adalah udara dengan permitivitas $\varepsilon_b = 1$. Panjang gelombang yang digunakan dalam hamburan satu silinder ini adalah $\lambda = 10^{-6}\ m$. Berdasarkan program yang telah dibuat maka perbandingan hasil hamburan untuk satu silinder dengan literatur ditunjukkan pada gambar 9 dan gambar 10.

Berdasarkan kedua gambar tersebut, maka program yang dibuat berhasil menentukan pola hamburan untuk satu silinder dan memiliki kesesuaian yang cukup baik dengan literatur referensi dalam penelitian ini. Dengan demikian dengan membuat suatu susunan periodik dari silinder-silinder tunggal tersebut maka dapat dibentuk susunan kristal fotonik 2D.

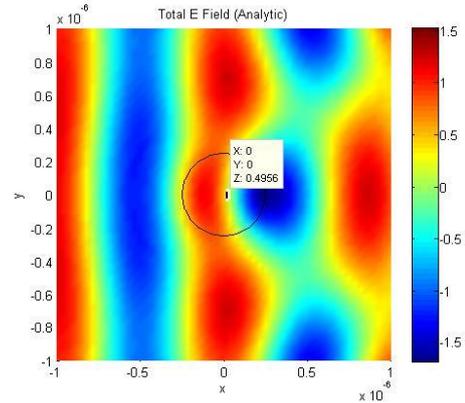

**Gambar 10** Hamburan medan EM TM pada satu silinder $\varepsilon_r = 4$ dan $\varepsilon_b = 1$ [10].

### Simulasi *Photonic bandgap* pada Kristal Fotonik 2D

Gambar 11 memperlihatkan bahwa terjadi PBG pada perambatan gelombang EM modus TM dalam kristal fotonik dengan silinder 6 x 6 yaitu tidak adanya gelombang EM yang ditansmisikan sepanjang kristal fotonik. Pada simulasi tersebut digunakan frekuensi sebesar $\frac{\omega a}{2\pi c} = 0,38$. Berdasarkan informasi dari gambar 9, maka frekuensi tersebut berada pada interval PBG. Permitivitas silinder yang digunakan adalah sebesar $\varepsilon_r = 11,56$.

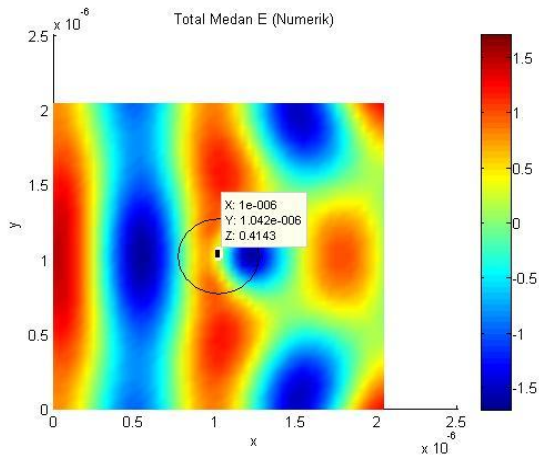

**Gambar 9**. Hamburan medan EM TM pada satu silinder dengan permitivitas silinder $\varepsilon_r = 4$ dan medium udara.

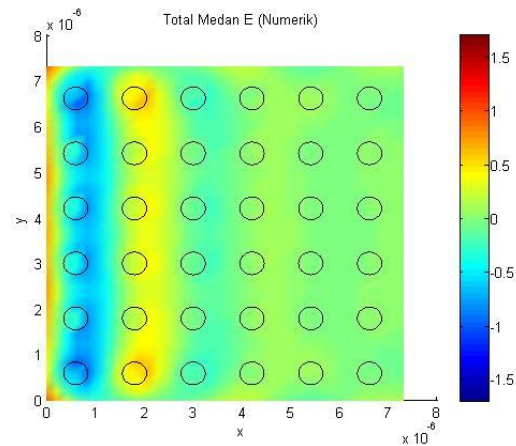

**Gambar 11**. *Photonic bandgap* pada Kristal fotonik 2D dengan parameter kisi $r = 0,2a$.

## Pandu Gelombang Kristal Fotonik 2D

Sebuah pandu gelombang di dalam susunan kristal fotonik 2D dibuat dengan mengambil salah satu susunan kolom atau baris silinder dalam struktur kristal.

Dalam penelitian ini sebuah kanal pandu gelombang dibuat dengan mengambil baris keempat dalam susunan kristal fotonik 2D berukuran 6 x 6. Dengan memilih salah satu frekuensi yang berada dalam *bandgap* maka didapatkan suatu bentuk pandu gelombang yang dapat mentransmisikan gelombang EM sepanjang kanal yang dibuat.

Pandu gelombang dalam kristal fotonik 2D ditunjukkan dalam gambar 12 dengan parameter jarak antar kisi $r = 0,2a$ dan permitivitas silinder $\varepsilon_r = 11,56$ dan frekuensi sebesar $\frac{\omega a}{2\pi c} = 0,38$.

Pada gambar tersebut dapat dilihat bahwa gelombang EM yang dirambatkan dalam kristal fotonik sesuai yang diperkirakan akan ditransmisikan melalui kanal dan terlihat bahwa medan listrik total yang merambat hanya berada pada kanal saja sedangkan pada daerah silinder di luar kanal besarnya medan listriknya mendekati nol. Bagian yang berwarna merah menunjukkan puncak gelombang sedangkan bagian yang berwarna biru menunjukkan lembah gelombang.

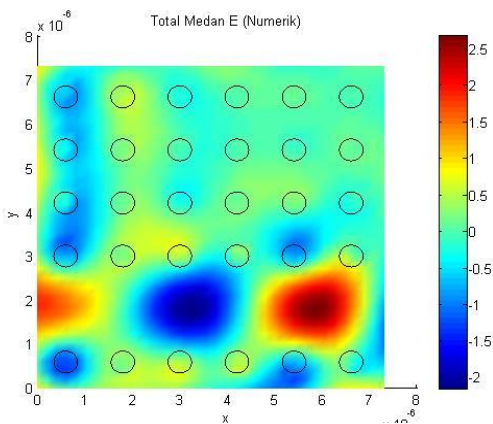

**Gambar 12** Perambatan gelombang EM TM dalam sebuah kanal pandu gelombang pada kristal fotonik 2D

## Kopling Medan EM pada Kristal Fotonik 2D dengan Dua Defek Simetris

Sebuah cacat (defek) adalah suatu struktur yang mengganggu keseimbangan struktural pada sebuah kristal. Dalam kristal fotonik 2D sebuah defek dapat berupa material yang berbeda indeks biasnya sebagai penyusun kristal, dan dapat pula dalam bentuk perbedaan ukuran silinder yang menyusun suatu kristal fotonik yang berasal dari material yang sama. Secara teoretis telah banyak dibuktikan bahwa kedua jenis defek tersebut memberikan manfaat dan kerugian tersendiri jika terdapat dalam susunan kristal fotonik 2D.

Dalam kasus penelitian ini kristal fotonik 2D yang digunakan akan diberikan dua cacat simetris disekitar kanal. Cacat yang diberikan pada kristal fotonik adalah dengan mengambil dua buah silinder yang berada pada baris keempat sehingga pada posisinya hanya terdapat medium saja berupa udara dengan permitivitas $\varepsilon_r = 1$.

Kristal fotonik 2D yang memiliki kanal dan dua cacat simetris tersebut memiliki parameter seperti pada kasus satu kanal sebelumnya dan diberikan variasi frekuensi gelombang datang. Variasi frekuensi gelombang datang yang digunakan adalah dari $\frac{\omega a}{2\pi c} = 0,36$ sampai $\frac{\omega a}{2\pi c} = 0,39$. Masing-masing frekuensi diberikan dalam bentuk simulasi sebagai berikut:

**Simulasi 1**

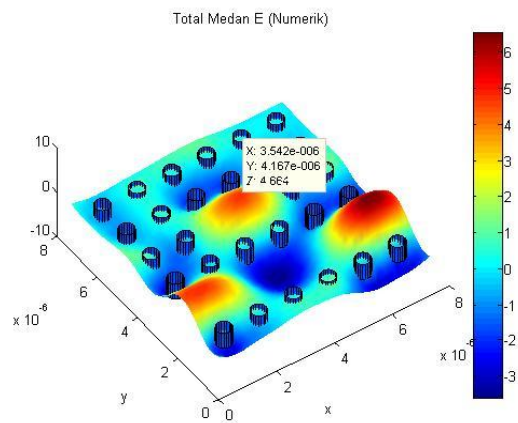

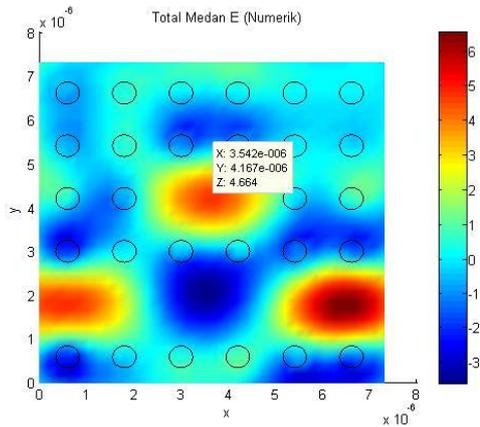

**Gambar 13** Simulasi kopling medan EM TM pada Kristal fotonik 2D dengan frekuensi $\frac{\omega a}{2\pi c} = 0,36$

Dari gambar 13 dapat dilihat bahwa terjadi kopling medan listrik yang terakumulasi dengan kanal yang ada dalam kristal fotonik 2D dengan amplitudo dari 0,2 hingga 4,6 sedangkan pada kanal amplitudo maksimum adalah 6,5.

**Simulasi 2**

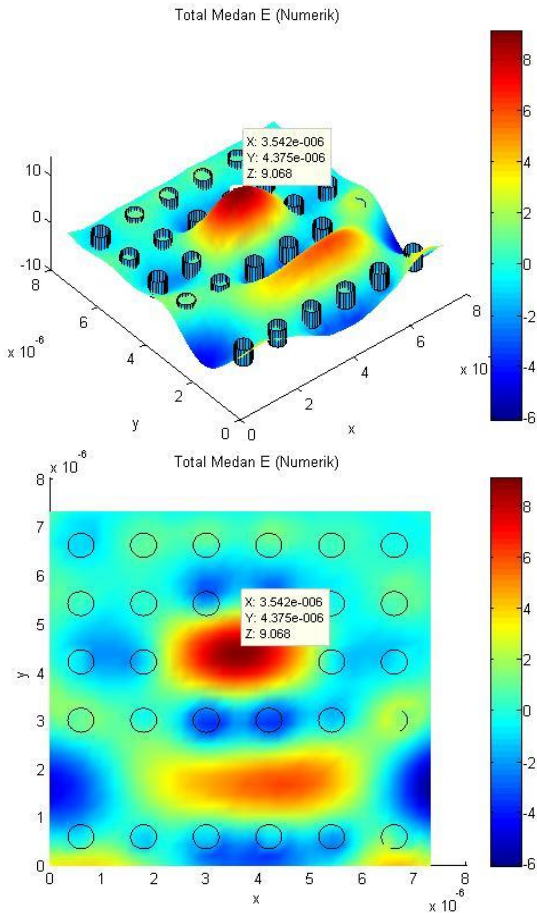

Dari gambar 14 di atas ditunjukkan bahwa pada frekuensi yang digunakan ternyata kopling medan EM yang dihasilkan memiliki amplitudo yang lebih besar daripada pada simulasi 1. Amplitudo yang dihasilkan pada model memiliki nilai yang berkisar dari 1,0 hingga 9,0 sedangkan amplitudo maksimum yang terjadi pada kanal adalah 6,1 sehingga model ini menghasillkan kopling medan EM yang lebih besar dibandingkan dengan model 1.

**Simulasi 3**

Gambar 15 memperlihatkan bahwa terdapat kopling medan EM pada frekuensi yang digunakan sebesar $\frac{\omega a}{2\pi c} = 0,38$. Amplitudo yang dihasilkan pada simulasi 3 ini menghasilkan nilai yang berkisar antara 0,5 hingga 10,6 sedangkan pada kanal nilai amplitudo maksimum yang dihasilkan adalah sebesar 8,2 seperti yang terlihat pada gambar 1 di atas.

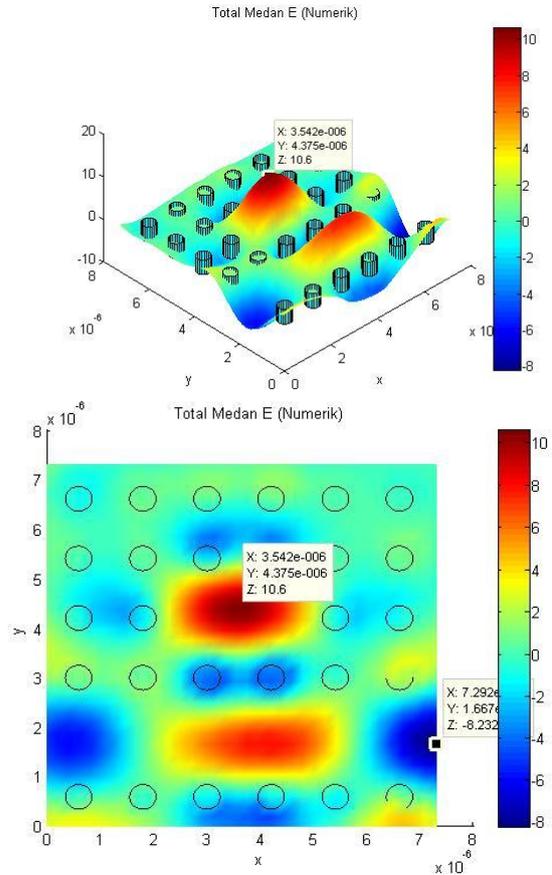

**Gambar 14** Simulasi kopling medan EM TM pada Kristal fotonik 2D dengan frekuensi $\frac{\omega a}{2\pi c} = 0,37$

**Gambar 15** Simulasi kopling medan EM TM pada Kristal fotonik 2D dengan frekuensi $\frac{\omega a}{2\pi c} = 0,38$

**Simulasi 4**

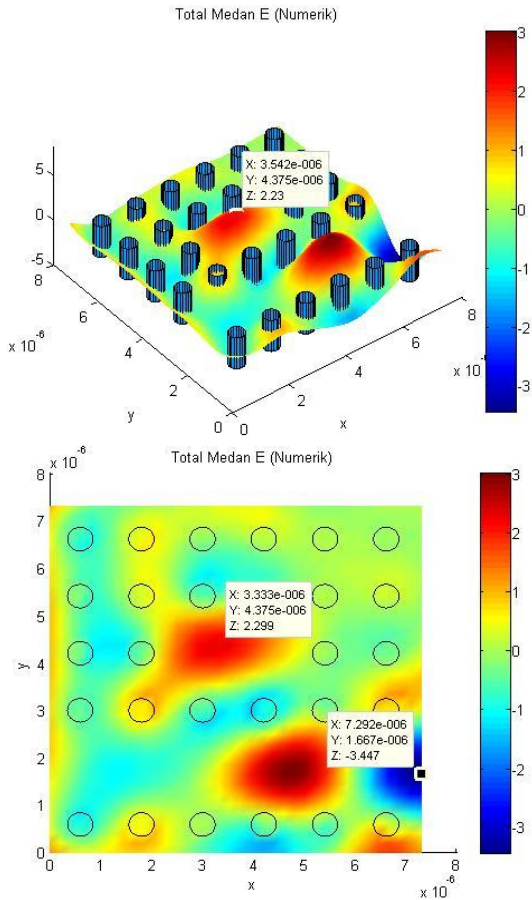

**Gambar 16** Simulasi kopling medan EM TM pada Kristal fotonik 2D dengan frekuensi $\frac{\omega a}{2\pi c} = 0,39$

Pada gambar 16 di atas terjadi kopling medan EM terhadap kanal seperti pada simulasi 1 sampai simulasi 3. Namun, pada simulasi 4 besarnya amplitudo pada defek berkurang besarnya dan berkisar pada nilai 0,05 hingga 2,23, sedangkan amplitudo maksimum pada kanal bernilai 3,4 yang berarti telah terjadi penurunan kembali tingkat kopling pada frekuensi ini dibandingkan model-model sebelumnya.

**Penghitungan Energi Medan EM dalam Daerah Cacat pada Kristal Fotonik 2D**

Berdasarkan beberapa model kristal fotonik dengan dua defek dan variasi frekuensi gelombang datang maka telah didapatkan terjadinya perubahan amplitudo yang terjadi pada peristiwa kopling medan EM antara defek dan kanal. Maka energi yang ada di dalam lokasi cacat dapat dihitung dan dibandingkan antara beberapa variasi frekuensi di sekitar *bandgap*.

Energi medan listrik pada perambatan gelombang EM didefinisikan sebagai:

$$W = \varepsilon \int_A \vec{E}_i \cdot \vec{E}_i \, dA_i \qquad (4.1)$$

dengan integral tersebut menyatakan jumlah keseluruhan kuadrat medan dalam luas area $A$. Luas area yang dimaksud adalah luas area cacat yaitu dua buah cacat titik yang ada dalam kristal fotonik yang diteliti. Luas area ini bersesuaian dengan kuadrat jarak antar kisi ($a$). Frekuensi yang diteliti adalah frekuensi yang digunakan dalam pemodelan sebelumnya, yaitu berkisar dari $\frac{\omega a}{2\pi c} = 0,36$ sampai $\frac{\omega a}{2\pi c} = 0,39$ dengan interval 0,01. Energi medan listrik pada area cacat ditunjukkan pada gambar,

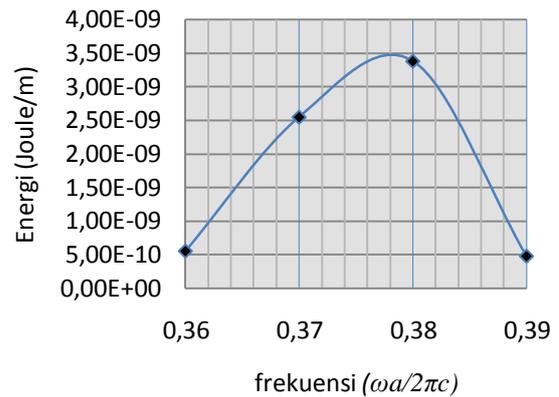

**Gambar 17** Daya medan listrik total pada area cacat dalam Kristal fotonik 2D dengan dua defek simetris

Besarnya energi medan listrik total pada cacat memiliki nilai maksimum. Frekuensi yang menghasilkan energi maksimum berada pada saat $\frac{\omega a}{2\pi c} = 0,378$ yang merupakan frekuensi efektif saat terjadi kopling pada daerah defek. Hal ini menandakan bahwa disekitar nilai tersebut maka besarnya energi medan akan berkurang sesuai dengan penurunan kurva. Saat frekuensi ditingkatkan atau dikurangi dari nilai frekuensi efektif maka besarnya kopling antara defek dan kanal juga akan semakin berkurang. Dengan demikian jika ingin memodifikasi besarnya energi yang terjadi pada defek maka cukup dengan mengubah-ubah frekuensinya.

Berdasarkan referensi nilai frekuensi efektif tersebut berada ditengah-tengah interval nilai *bandgap* seperti yang ditunjukkan pada gambar 9. Besarnya frekuensi efektif untuk menghasilkan energi maksimum pada defek akan tepat di tengah-tengah interval *bandgap* jika struktur *mesh* yang dibuat lebih besar lagi (sehingga akan butuh komputer dengan memori yang lebih besar juga) [2].

Secara umum jika lokasi defek dipindahkan lebih jauh dari kanal dengan jarak 3 kali jarak antar kisi maka frekuensi efektif yang berlaku pada model tersebut akan sama namun hasil energi medan total akan berkurang sebanding dengan pertambahan jarak antara defek dan kanal.

# KESIMPULAN DAN SARAN

**Kesimpulan**

Metode tensor green dapat digunakan untuk menghitung kuat medan listrik total dalam suatu krital fotonik 2D. Dengan bantuan program MATLAB maka simulasi penghitungan kuat medan listrik total pada kristal fotonik dengan metode tensor green dapat dilakukan.

Dalam penelitian ini diberikan perlakuan pada kristal fotonik 2D dengan adanya sebuah kanal pandu gelombang dan dua buah defek simetris yang disusun berhimpitan. Dari data-data yang didapat dapat disimpulkan bahwa pada struktur kristal fotonik 2D tersebut terjadi peristiwa kopling antara kanal dengan defek yang disusun simetris terhadap kanal. Besarnya amplitudo dari medan listrik dalam defek bergantung pada frekuensi gelombang datang pada kristal fotonik 2D. Dengan adanya peristiwa kopling tersebut maka besar kuat medan listrik yang merambat dalam kanal akan terpecah saat terjadi kopling dengan sebagian medan tersebut ditransmisikan keluar dari struktur kristal fotonik 2D.

Melalui analisis grafik *bandgap* dari literatur terlihat bahwa frekuensi efektif yang menghasilkan energi maksimum pada saat terjadi kopling berada di pertengahan selang frekuensi *bandgap*.

**Saran**

Untuk mengdapatkan hasil kopling yang lebih maksimal dapat digunakan jumlah silinder (*rod*) yang lebih banyak karena dapat mengurangi "kebocoran" medan disekitar struktur kristal fotonik 2D. Dengan adanya peristiwa kopling dalam penelitian ini maka pengujian menggunakan jumlah defek yang lebih banyak dan berada pada jarak tertentu sehingga dimungkinkan akan bermanfaat dalam pembuatan *channel drop filter* (dengan menambah jumlah memori komputer yang digunakan tentunya).

# Lampiran

# Lampiran 1

**Penurunan bentuk vektor fungsi Green 3D**

Medan listrik $\vec{E}(r)$ dan rapat arus $\vec{J}(r)$ dihubungkan melalui persamaan :

$$\vec{E}(\vec{r}) = i\omega\mu \int_V \vec{G}(\vec{r},\vec{r}') \cdot \vec{J}(\vec{r}') d\vec{r}' \qquad (1)$$

yang merupakan solusi dari persamaan :

$$\vec{\nabla} \times \vec{\nabla} \times \vec{E}(\vec{r}) - k_0^2 \vec{E}(\vec{r}) = i\omega\mu\vec{J}(\vec{r}) \qquad (2)$$

Subtitusikan persamaan (1) ke (2) :

$$\vec{\nabla} \times \vec{\nabla} \times \vec{G}(\vec{r},\vec{r}') - k_0^2 \vec{G}(\vec{r},\vec{r}') = \bar{I}\delta(\vec{r}-\vec{r}') \qquad (3)$$

dengan $\bar{I}$ disebut sebagai unit dyad. Medan listrik adalah $\vec{E}(\vec{r})$ solusi dari persamaan Maxwell oleh keberadaan sumber $\vec{J}(\vec{r})$

$$\vec{\nabla} \times \vec{E}(\vec{r}) = i\omega\mu\vec{H}(\vec{r}) \qquad (4)$$

$$\vec{\nabla} \times \vec{H}(\vec{r}) = -i\omega\varepsilon\vec{E}(\vec{r}) + \vec{J}(\vec{r}) \qquad (5)$$

Karena $\nabla \cdot \mu H(r) = 0$ sehingga $\mu H(r)$ harus memiliki bentuk gausian $\mu H(r) = \nabla \times A(r)$ dengan $\vec{A}$ adalah potensial vektor. Subtitusikan ke dalam persamaan (4) :

$$\vec{\nabla} \times \vec{E}(\vec{r}) = i\omega\vec{\nabla} \times \vec{A}(\vec{r}) \qquad (6)$$

karena $\vec{\nabla} \times \vec{\nabla}\phi = 0$ maka persamaan (6) menjadi :

$$\vec{E}(\vec{r}) = i\omega\vec{A}(\vec{r}) - \vec{\nabla}\phi(\vec{r}) \qquad (7)$$

dengan $\nabla\phi$ adalah potensial skalar. Maka persamaan Maxwell kedua menjadi :

$$\vec{\nabla} \times \vec{\nabla} \times \vec{A}(\vec{r}) = \omega^2\mu\varepsilon\vec{A}(\vec{r}) + i\omega\mu\varepsilon\vec{\nabla}\phi(\vec{r}) + \mu\vec{J}(\vec{r}) \qquad (8)$$

Dengan menggunakan identitas vektor :

$$\vec{\nabla} \times \vec{\nabla} \times \vec{A} = \vec{\nabla}(\vec{\nabla} \cdot \vec{A}) - \vec{\nabla}^2\vec{A} \qquad (9)$$

Persamaan (8) menjadi berbentuk:

$$\vec{\nabla}^2\vec{A}(\vec{r}) + k_0^2\vec{A}(\vec{r}) - \vec{\nabla}[\vec{\nabla} \cdot \vec{A}(\vec{r})] = -i\omega\mu\varepsilon\vec{\nabla}\phi(\vec{r}) - \mu\vec{J}(\vec{r}) \qquad (10)$$

Gunakan gauge Lorentz:

$$\vec{\nabla} \cdot \vec{A}(\vec{r}) = +i\omega\mu\varepsilon\phi(\vec{r}) \qquad (11)$$

Pada persamaan (10) sehingga didapat :

$$(\vec{\nabla}^2 + k_0^2)\vec{A}(\vec{r}) = -\mu\vec{J}(\vec{r}) \tag{12}$$

pada media yang tak homogen, $\vec{\nabla}\cdot\vec{E} = \rho/\varepsilon$ kemuian dengan mengambil divergensinya dan menggunakan gauge Lorentz didapatkan:

$$\vec{E}(\vec{r}) = i\omega\vec{A}(\vec{r}) - \vec{\nabla}\phi(\vec{r}) \tag{13}$$

maka

$$(\vec{\nabla}^2 + k_0^2)\phi(\vec{r}) = -\rho(\vec{r})/\varepsilon \tag{14}$$

Dengan menggunakan fungsi Green skalar pada:

$$(\vec{\nabla}^2 + k_0^2)\vec{A}(\vec{r}) = -\mu\vec{J}(\vec{r}) \tag{15}$$

sehingga

$$\vec{A}(\vec{r}) = \mu\int_V d\vec{r}'\, g(\vec{r},\vec{r}')\vec{J}(\vec{r}') \tag{16}$$

Dengan menggunakan cara yang sama, solusi dari :

$$(\vec{\nabla}^2 + k_0^2)\phi(\vec{r}) = -\rho(\vec{r})/\varepsilon \tag{17}$$

adalah

$$\phi(\vec{r}) = \frac{1}{\varepsilon}\int_V d\vec{r}'\, g(\vec{r},\vec{r}')\rho(\vec{r}') \tag{18}$$

Sehingga medan listriknya

$$\vec{E}(\vec{r}) = i\omega\vec{A}(\vec{r}) - \vec{\nabla}\phi(\vec{r}) \tag{19}$$

ditunjukkan oleh:

$$\vec{E}(\vec{r}) = i\omega\mu\int_V d\vec{r}'\, g(\vec{r},\vec{r}')J(\vec{r}') - \frac{\nabla}{\varepsilon}\int_V d\vec{r}'\, g(\vec{r},r')\rho(\vec{r}') \tag{20}$$

Dari relasi kontinu,

$$\vec{E}(\vec{r}) = i\omega\mu\int_V d\vec{r}'\, g(\vec{r},\vec{r}')J(\vec{r}') - \frac{\vec{\nabla}\vec{\nabla}\cdot}{i\omega\varepsilon}\int_V d\vec{r}'\, g(\vec{r},\vec{r}')\vec{J}(\vec{r}') \tag{21}$$

Dengan melakukan integrasi parsial pada suku kedua dari persamaan (21) dan menggunakan sifat simetrik $g(r,r')$ mengenai subtitusi pada $\vec{r}$ dan $\vec{r}'$ yaitu :

$$\vec{E}(\vec{r}) = i\omega\mu\int_V d\vec{r}'\, \bar{\vec{G}}(\vec{r},\vec{r}')\vec{J}(\vec{r}') \tag{22}$$

maka $\overline{G}(r,r')$ menjadi berbentuk :

$$\vec{\vec{G}}(\vec{r},\vec{r}\,') = \left[\vec{I} + \frac{\vec{\nabla}\vec{\nabla}}{k_B^2}\right] g(\vec{r},\vec{r}\,') \qquad (23)$$

yang disebut sebagai tensor Green 3D.